%% file: eprint.tex
\documentclass[12pt]{article}
\usepackage{graphicx,amsmath,amsthm,amssymb,cite}

\DeclareMathOperator{\diag}{diag}
\newcommand{\eq}{Eq.}

\newcommand\pubnumber{SNSN-323-63}
\newcommand\pubdate{\today}

\def\ift{Instituto de F\'isica Te\'orica UAM/CSIC,
 Calle Nicol\'as Cabrera 13-15, Cantoblanco E-28049 Madrid, Spain\\}
\def\uam{Departamento de F\'isica Te\'orica, Universidad Aut\'onoma de Madrid, Cantoblanco E-28049 Madrid, Spain}
\def\cern{CERN, Theoretical Physics Department, Geneva, Switzerland.}
\def\supportuno{\footnote{Work supported by EU through grants H2020-MSCA-ITN-2015/674896-Elusives and H2020-MSCA-RISE-2015/690575-InvisiblesPlus and by the EU FP7 Marie Curie Actions CIG NeuProbes (PCIG11-GA-2012-321582) and the Spanish MINECO through the ``Ram\'on y Cajal'' programme (RYC2011-07710), the project FPA2012-31880,  through the Centro de Excelencia Severo Ochoa Program under grant SEV-2012-0249.}}
\def\supportdos{\footnote{Work supported by EU through grants H2020-MSCA-ITN-2015/674896-Elusives and H2020-MSCA-RISE-2015/690575-InvisiblesPlus.}}

\def\Title#1{\begin{center} {\Large #1 } \end{center}}
\def\Author#1{\begin{center}{ \sc #1} \end{center}}
\def\Address#1{\begin{center}{ \it #1} \end{center}}

\newcommand\pubblock{\rightline{\begin{tabular}{l} \pubnumber\\
         \pubdate  \end{tabular}}}
\newenvironment{Abstract}{\begin{quotation}  }{\end{quotation}}
\newenvironment{Presented}{\begin{quotation} \begin{center} 
             PRESENTED AT\end{center}\bigskip 
      \begin{center}\begin{large}}{\end{large}\end{center} \end{quotation}}
\def\Acknowledgements{\bigskip  \bigskip \begin{center} \begin{large}
             \bf ACKNOWLEDGEMENTS \end{large}\end{center}}
\input econfmacros.tex
\begin{document}
\begin{titlepage}
\pubblock

\vfill
\Title{Non-Unitarity vs sterile neutrinos at DUNE}
\vfill
\Author{Josu Hernandez-Garcia\supportuno}
\Address{\ift \uam}
\Author{Jacobo Lopez-Pavon\supportdos}
\Address{\cern}

\vfill
\begin{Abstract}
Neutrino masses are one of the most promising open windows to physics beyond the Standard Model (SM). Several extensions of the SM which accommodate neutrino masses require the addition of right-handed neutrinos to 
its particle content. These extra fermions will either be kinematically accessible (sterile neutrinos) or not (deviations from Unitarity of the PMNS matrix) but at some point they will impact neutrino oscillation searches. 
We explore the differences and similitudes between the two cases and compare their present bounds with the 
expected sensitivities of DUNE. We conclude that Non-Unitarity (NU) effects are too constrained to impact present 
or near future neutrino oscillation facilities but that sterile neutrinos can play an important role at long 
baseline experiments.
\end{Abstract}
\vfill
\begin{Presented}
NuPhys2016, Prospects in Neutrino Physics

Barbican Centre, London, UK,  December 12--14, 2016
\end{Presented}
\vfill
\end{titlepage}
\def\thefootnote{\fnsymbol{footnote}}
\setcounter{footnote}{0}

\section{Introduction}

The simplest extension of the Standard Model (SM) of particle physics able to account for the evidence for neutrino masses 
and mixings consists in the addition of right-handed neutrinos to its particle content. The new physics scale is 
the Majorana mass of the new states and, since it is not related to the electroweak symmetry breaking mechanism, there is no 
theoretical guidance for its value. A large Majorana scale leads to the celebrated seesaw mechanism~\cite{Minkowski:1977sc,Mohapatra:1979ia,Yanagida:1979as,GellMann:1980vs}, providing a very natural explanation of the lightness of neutrino masses. Conversely, a light neutrino mass could also naturally stem from a symmetry argument~\cite{Mohapatra:1986bd,Bernabeu:1987gr,Branco:1988ex,Buchmuller:1990du,Pilaftsis:1991ug,Kersten:2007vk,Abada:2007ux}. 
This proceeding is based on~\cite{Blennow:2016jkn} where we analyze the phenomenological impact of these new physics in 
neutrino oscillation facilities. If the new mass scale is kinematically accessible in meson decays, the sterile states 
will be produced in the neutrino beam. On the other hand, if the extra neutrinos are too heavy to be produced, the 
effective $3\times 3$ PMNS matrix will show unitarity deviations. We will refer to these situations as sterile and 
Non-Unitary (NU) neutrino oscillations, respectively. The aim of our work is to discuss the similitudes and differences among
these two regimes clarifying in which limit they lead to the same neutrino oscillation phenomenology.

\section{Non-unitarity and sterile $\nu$ phenomenology}
\label{sec:comparison}

If $n$ extra right-handed neutrinos are added to the SM Lagrangian, the full mixing matrix (including both light and heavy states) can be written as
\begin{equation}
\label{eq:Ufull}
\mathcal{U} = \begin{pmatrix}
N & \Theta  \\
R & S \\
\end{pmatrix} \, ,
\end{equation}
where $N$ represents the $3 \times 3$ active-light sub-block (i.e., the PMNS matrix), which will no longer be unitary. Here, $\Theta$ is the $3\times n$ sub-block that includes the mixing between active and heavy states, while the $R$ and $S$ sub-blocks define the mixing of the sterile states with the light and heavy states, respectively. Note that both $R$ and $S$ are not involved when considering oscillations among active flavours.

\subsection{Non-unitarity case}
In the case of NU, only the light states are kinematically accessible and the amplitude for producing one of these states and a charged lepton of flavour $\alpha$ in a particular decay is proportional to the mixing matrix element $N_{\alpha i}^*$. In the mass eigenstate basis, the evolution of the produced neutrino state is given by the Hamiltonian~\cite{Antusch:2006vwa}
\begin{equation}
\label{eq:Hnonunitarity}
H = \frac{1}{2E}\begin{pmatrix}
0 & 0 & 0 \\
0 & \Delta m_{21}^2 & 0 \\
0 & 0 & \Delta m_{31}^2
\end{pmatrix} + N^\dagger  \begin{pmatrix}
V_{\rm CC}+V_{\rm NC} & 0 & 0 \\
0 & V_{\rm NC} & 0 \\
0 & 0 & V_{\rm NC}
\end{pmatrix}  N,
\end{equation}
where $V_{\rm CC}=\sqrt{2}G_Fn_e$ and $V_{\rm NC}=-G_F n_n/\sqrt{2}$ are the charged-current (CC) and neutral-current (NC) matter potentials, respectively. The amplitude for a neutrino in the mass eigenstate $j$ to interact as a neutrino of flavour $\beta$ is given by the mixing 
matrix element $N_{\beta j}$, which means that the oscillation probability will be given by
\begin{equation}
\label{eq:nonunitarityoscillationformula}
P_{\alpha\beta} = |(N S^0 N^\dagger)_{\beta\alpha}|^2,
\end{equation}
where $S^0 = \exp(- i H L)$ for a constant matter potential. Here $P_{\alpha\beta}$ denotes the ``theoretical'' oscillation probabilities, defined as the ratio between the observed number of events divided by the product of the SM-predicted flux times cross section. However, in practice neutrino oscillation experiments do not measure $P_{\alpha\beta}$. Most present and future experiments rather determine the flux and cross sections via near detector data and extrapolate to the far detector by correcting for the different geometries, angular apertures, and detection cross sections. For the near detector, we assume that the phases corresponding to the propagation of the light neutrinos have not yet developed significantly and therefore $S^0 = I$, resulting in the experimentally inferred probability
\begin{equation}
\label{eq:experimentalprobability}
\mathcal P_{\alpha\beta} = \frac{\left| (N S^0 N^\dagger)_{\beta\alpha} \right|^2}{((NN^\dagger)_{\alpha\alpha})^2}.
\end{equation}
In the SM limit the matrix $N$ becomes unitary and $\mathcal P_{\alpha\beta} = P_{\alpha\beta}$ as expected.

\subsection{Sterile neutrino case}

In the sterile neutrino scenario, all of the states are kinematically accessible and the oscillation evolution matrix $\mathcal S$, involving both light and heavy states, takes the form
\begin{equation}
\mathcal S = \mathcal U \mathcal S^0 \mathcal U^\dagger,
\end{equation}
where $\mathcal S^0$ is the full $(3+n)\times (3+n)$ evolution matrix expressed in the mass eigenbasis. For vacuum oscillations, we find that $\mathcal S^0 = \diag(\exp(-i\Delta m_{j1}^2L/2E))$. Therefore, the active neutrino $3\times 3$ sub-block $S$ can be simplified to 
\begin{equation}
S_{\alpha\beta} = \sum_{i \in \rm light}  N_{\alpha i} S^0_{ij} N^*_{\beta j} + \sum_{J \in \rm heavy} \Theta_{\alpha J}\Theta_{\beta J}^* \Phi_J,
\end{equation}
where $\alpha,\beta$ stand for active neutrino flavors, $\Phi_J$ is the phase factor acquired by the heavy state $J$ as it propagates, and $S^0$ is defined in the same way as in the NU case. In the limit $\Delta m^2_{iJ} L/E \gg 1$, the oscillations are too fast to be resolved at the detector and only an averaged-out effect is observable. In this averaged-out limit, the cross terms between the first and second term average to zero and we find
\begin{equation}
\label{eq:sterileoscillationformula}
P_{\alpha\beta} = |S_{\alpha\beta}|^2 = \left| \sum_i N_{\alpha i} S^0_{ij} N^*_{\beta j} \right|^2 + \mathcal O(\Theta^4) \, ,
\end{equation}
which recovers the same expression as in the NU case given in \eq~\eqref{eq:nonunitarityoscillationformula}. 
%Thus, we can conclude that averaged-out sterile neutrino oscillations in vacuum are equivalent to NU to 
%leading order. 
For oscillations in the presence of matter, the sterile neutrino oscillations will be subjected to a matter potential
\begin{equation}
\mathcal H_{\rm mat}^f = \begin{pmatrix}
V_{3 \times 3} & 0 \\ 0 & 0
\end{pmatrix} \quad \text{with} \quad V_{3\times 3} = \begin{pmatrix}
V_{\rm CC}+V_{\rm NC} & 0 & 0 \\
0 & V_{\rm NC} & 0 \\
0 & 0 & V_{\rm NC}
\end{pmatrix} \, .
\end{equation}
If the matter potential is small in comparison to the light-heavy energy splitting $\Delta m_{iJ}^2/2E$, the light-heavy 
mixing in matter will be given by
\begin{equation}
\tilde{\Theta}_{\alpha J} = \Theta_{\alpha J} + \frac{2E}{\Delta m_{iJ}^2}(\delta_{\alpha e} V_{\rm CC} \Theta_{eJ} + \Theta_{\alpha J} V_{\rm NC})
\end{equation}
to first order in perturbation theory. In the limit $\Delta m_{iJ}^2/2E \gg V_{\rm CC}, V_{\rm NC}$, we canntherefore neglect the difference between the heavy mass eigenstates in vacuum and in matter, and apply Eq.~\eqref{eq:sterileoscillationformula}. 
Thus, the matter Hamiltonian in the light sector can be computed in exactly the same way as for the NU scenario and we 
find again the same expressions for the theoretical probability in Eq.~\eqref{eq:nonunitarityoscillationformula} as for the NU case, 
at leading order in $\Theta$. 

Also in the sterile neutrino case the impact of the near detector measurements on the extraction 
of the experimentally measurable probability should be taken into account. In this work we will always assume that the oscillations due to the additional 
heavy states are averaged out at the far detector. However, this might not be the case at the near detector. We will focus on the following two scenarios:
\begin{enumerate}
\item The light-heavy oscillations are averaged out already at the near detector. For practical purposes, the oscillation 
phenomenology in this case is identical to the NU case and \eq~\eqref{eq:experimentalprobability} also applies. For the 
experimental setup of DUNE, with a peak neutrino energy of $\sim 2.5$~GeV and a near detector distance of $\sim0.5$~km, 
this is the case when $\Delta m^2 \gtrsim 100~\textrm{eV}^2$.
\item The light-heavy oscillations have not yet developed at the near detector, but are averaged out at the far detector. In 
this case, the near detector would measure the SM fluxes and cross sections, and therefore the denominator in 
Eq.~\eqref{eq:experimentalprobability} would be equal to one. In this case, the experimental probability would 
coincide with the ``theoretical'' probability in Eq.~\eqref{eq:nonunitarityoscillationformula}. This scenario is the one 
implicitly assumed in many phenomenological studies. %, given the simplicity of Eq.~\eqref{eq:nonunitarityoscillationformula}. 
For DUNE, since the far detector baseline is $1300$~km, this would be the case only in the region 
$0.1~\textrm{eV}^2 \lesssim \Delta m^2 < 1~\textrm{eV}^2$. 
\end{enumerate}

%%%%%%%%%%%%%%%%%%%
\section{Parametrizations}
\label{sec:param}
%%%%%%%%%%%%%%%%%%%

The non-unitarity effects stemming from the heavy-active neutrino mixing can be parameterized as 
\begin{equation}
N = T U = (I-\alpha) U \quad \text{where} \quad \alpha = (1-T) = \begin{pmatrix}
\alpha_{ee} & 0 & 0\\
\alpha_{\mu e} & \alpha_{\mu \mu} & 0\\
\alpha_{\tau e} & \alpha_{\tau \mu} & \alpha_{\tau \tau}
\end{pmatrix}\, ,
\label{eq:N}
\end{equation}
where $T$ is a lower triangular matrix~\cite{Xing:2007zj,Xing:2011ur,Escrihuela:2015wra} and $\alpha_{\alpha \beta}\ll1$. In Eq.~\eqref{eq:N} $U$ is 
a unitary matrix that is equivalent to the standard PMNS matrix up to small corrections proportional to the deviations 
encoded in $\alpha$. When the normalization at the near detector is considered, it effectively cancels any leading 
order dependence on $\alpha_{\beta\beta}$ in disappearance channels in vacuum. Expanding in $\alpha_{\delta\gamma}$ we 
obtain
\begin{equation}
\mathcal P_{\alpha\beta}\simeq\left|\left(1+\alpha_{\alpha\alpha}-\alpha_{\beta\beta}\right)(U S^0 U^\dagger)_{\alpha\beta}  
-\sum_{\delta\neq\alpha}\alpha_{\alpha\delta}(U S^0 U^\dagger)_{\delta\beta}-
\sum_{\delta\neq\beta}\alpha^*_{\beta\delta}(U S^0 U^\dagger)_{\alpha\delta}\right|^2
\label{eq:cancellation}
\end{equation}
When $\alpha=\beta$ the dependence on $\alpha_{\beta\beta}$ cancels out. This shows how relevant the role of the near 
detectors is regarding the sensitivity to the new physics parameters.

An alternative and widely used parameterization is $N=(1-\eta)U^{'}$ where $\eta=\eta^\dagger$ and $U^{'}$ is a unitary matrix~\cite{Broncano:2002rw,FernandezMartinez:2007ms}. The mapping between this hermitian parameterization and the triangular
parametrization given in Eq.~\eqref{eq:N} is provided in\cite{Blennow:2016jkn}.

%%%%%%%%%%%%%%%%%%%%%%%%%%%%%%%
\section{Present constraints on deviations from unitarity}
\label{sec:bounds}
%%%%%%%%%%%%%%%%%%%%%%%%%%%%%%%
PMNS NU from very heavy extra neutrinos modifies precision electroweak and flavour observables 
(see for instance ~\cite{Shrock:1980vy,Antusch:2006vwa,Antusch:2014woa,Fernandez-Martinez:2016lgt}). 
These modification translate into very strong upper limits on the $\alpha$ parameters taken from 
Ref.~\cite{Fernandez-Martinez:2016lgt} and listed in the left column in Table~\ref{tab:bounds}. However, for sterile 
neutrinos with masses below the electroweak scale these stringent constraints are lost, since all mass eigenstates 
are kinematically available in the observables used to derive the constraints and unitarity is therefore restored. 
The sensitivity of oscillation experiments to sterile neutrino mixing will depend on the actual value of the sterile 
neutrino mass, which determines if the corresponding $\Delta m^2$ leads to oscillations for the energy and baseline that 
characterize the experimental setup. Once oscillations are averaged-out, the constraints derived will become independent 
of $\Delta m^2$. The bounds in the middle column apply for $\Delta m^2 \gtrsim 100$~eV$^2$ and will thus be relevant when 
the sterile neutrino oscillations are in the averaged-out regimes for both the near and far detectors of most long-baseline 
experiments. The bounds in the right column apply for $\Delta m^2 \sim 0.1-1$~eV$^2$ and will thus be relevant when the 
sterile neutrino oscillations are in the averaged-out regime for the far detector, but not for the near detector. For a 
more comprehensive breakdown of the available constraints and their ranges of applicability, we refer the interested reader 
to Appendix A of~\cite{Blennow:2016jkn}.

%%%%%%%%%%%%%%%%%%%%%%
\begin{table}[t]
\setlength{\tabcolsep}{7pt}
\begin{center}
\renewcommand{\arraystretch}{1.6}
\begin{tabular}{|  c@{\quad} | c@{\quad} | c@{\quad} c@{\quad}   | }
\hline
    & ``Non-Unitarity'' & \multicolumn{2}{c|}{``Light steriles''} \\
     & ($m>$ EW)  &  $\Delta m^2 \gtrsim 100$~eV$^2$ & $\Delta m^2 \sim 0.1-1$~eV$^2$\\ \hline\hline
$\alpha_{ee} $ & $1.3 \cdot 10^{-3}$~\cite{Fernandez-Martinez:2016lgt} & $2.4 \cdot 10^{-2} $~\cite{Declais:1994su} & $1.0 \cdot 10^{-2} $~\cite{Declais:1994su}\\
$\alpha_{\mu\mu}$ & $2.2 \cdot 10^{-4}$~\cite{Fernandez-Martinez:2016lgt} & $2.2 \cdot 10^{-2}$~\cite{Abe:2014gda} & $1.4 \cdot 10^{-2}$~\cite{MINOS:2016viw}\\
$\alpha_{\tau\tau}$ & $2.8 \cdot 10^{-3}$~\cite{Fernandez-Martinez:2016lgt} & $1.0 \cdot 10^{-1}$~\cite{Abe:2014gda} & $1.0 \cdot 10^{-1}$~\cite{Abe:2014gda}\\
$\lvert\alpha_{\mu e}\rvert$ & $6.8 \cdot 10^{-4} \; (2.4 \cdot 10^{-5})$~\cite{Fernandez-Martinez:2016lgt} & $2.5 \cdot 10^{-2} $~\cite{Astier:2003gs} & $1.7 \cdot 10^{-2} $ \\
$\lvert\alpha_{\tau e}\rvert$ & $2.7 \cdot 10^{-3}$~\cite{Fernandez-Martinez:2016lgt} & $6.9 \cdot 10^{-2}$ & $4.5 \cdot 10^{-2}$ \\
$\lvert\alpha_{\tau\mu}\rvert$ & $1.2 \cdot 10^{-3}$~\cite{Fernandez-Martinez:2016lgt} & $1.2 \cdot 10^{-2}$~\cite{Astier:2001yj} & $5.3 \cdot 10^{-2}$ \\ \hline\hline
\end{tabular}
\caption{\label{tab:bounds} Current upper bounds on the $\alpha$ parameters. The limits are shown 
at $2\sigma$ for the NU case and 95\% CL (1 d.o.f.) for the light sterile neutrino limit.}
\end{center}
\end{table}

%%%%%%%%%%%%%%%%%%%
\section{DUNE sensitivities}
\label{sec:simulations}
%%%%%%%%%%%%%%%%%%%

The choice of the facility under 
study is motivated by the strong matter effects that characterize the DUNE setup and that allow to probe not only the source 
and detector effects induced by the new physics in a given channel $P_{\alpha\beta}$, but also the matter effects which 
now provide sensitivity to other $\alpha_{\gamma \rho}$ parameters not necessarily satisfying 
$\gamma, \rho \leq \alpha$ or $\gamma, \rho \leq \beta$. In the fit, the assumed true values for the standard oscillation 
parameters are set according to their current best-fits from Ref.~\cite{Gonzalez-Garcia:2014bfa}. The mixing angles and 
squared-mass splittings are allowed to vary in the simulations, using a Gaussian prior corresponding to their current 
experimental uncertainties from Ref.~\cite{Gonzalez-Garcia:2014bfa} centered around their true values. The CP-violating 
phase is left completely free during the simulations, and its true value is set to $\delta_\mathrm{CP}=-\pi/2$. We have 
performed simulations for two distinct new physics scenarios. In the first case (ND averaged) we normalize the oscillation 
probabilities according to Eq.~\eqref{eq:experimentalprobability}. For the DUNE setup, the requirement for having 
averaged-out oscillations at the near detector translates to the condition $\Delta m^2 \gtrsim $~100 eV$^2$. The second 
scenario ({ND undeveloped}) would correspond to the case where sterile neutrino oscillations are averaged out at the 
far detector but have not yet developed at the near detector (this is 
$0.1~\text{eV}^2 \lesssim \Delta m^2 < 1~\text{eV}^2$ for the DUNE setup). In this case, no extra normalization is needed and the 
oscillation probability is directly given by Eq.~\eqref{eq:nonunitarityoscillationformula}.
\begin{figure}[t]
 \includegraphics[width=0.48\columnwidth]{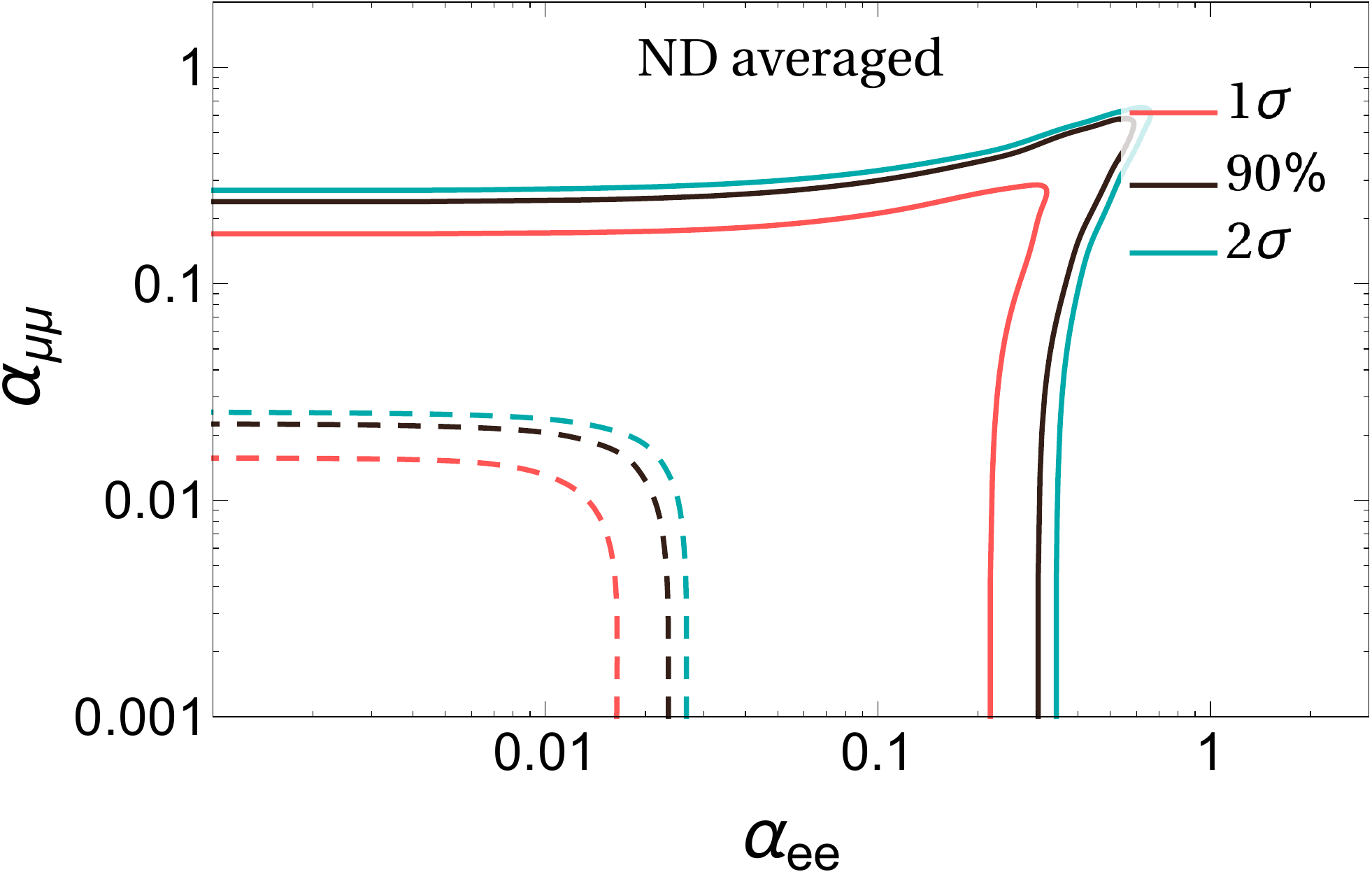} 
 \includegraphics[width=0.48\columnwidth]{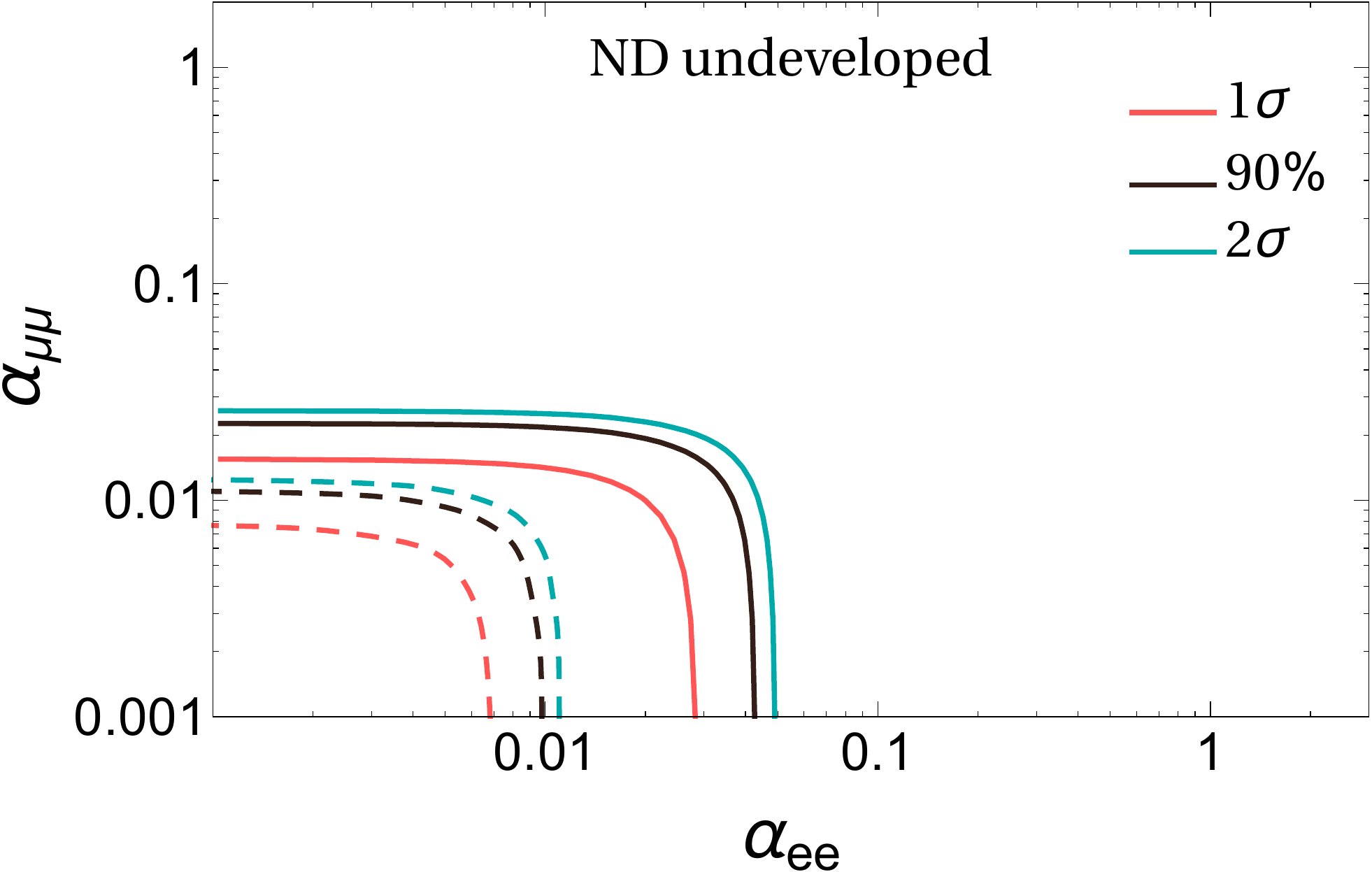} \\
  \includegraphics[width=0.48\columnwidth]{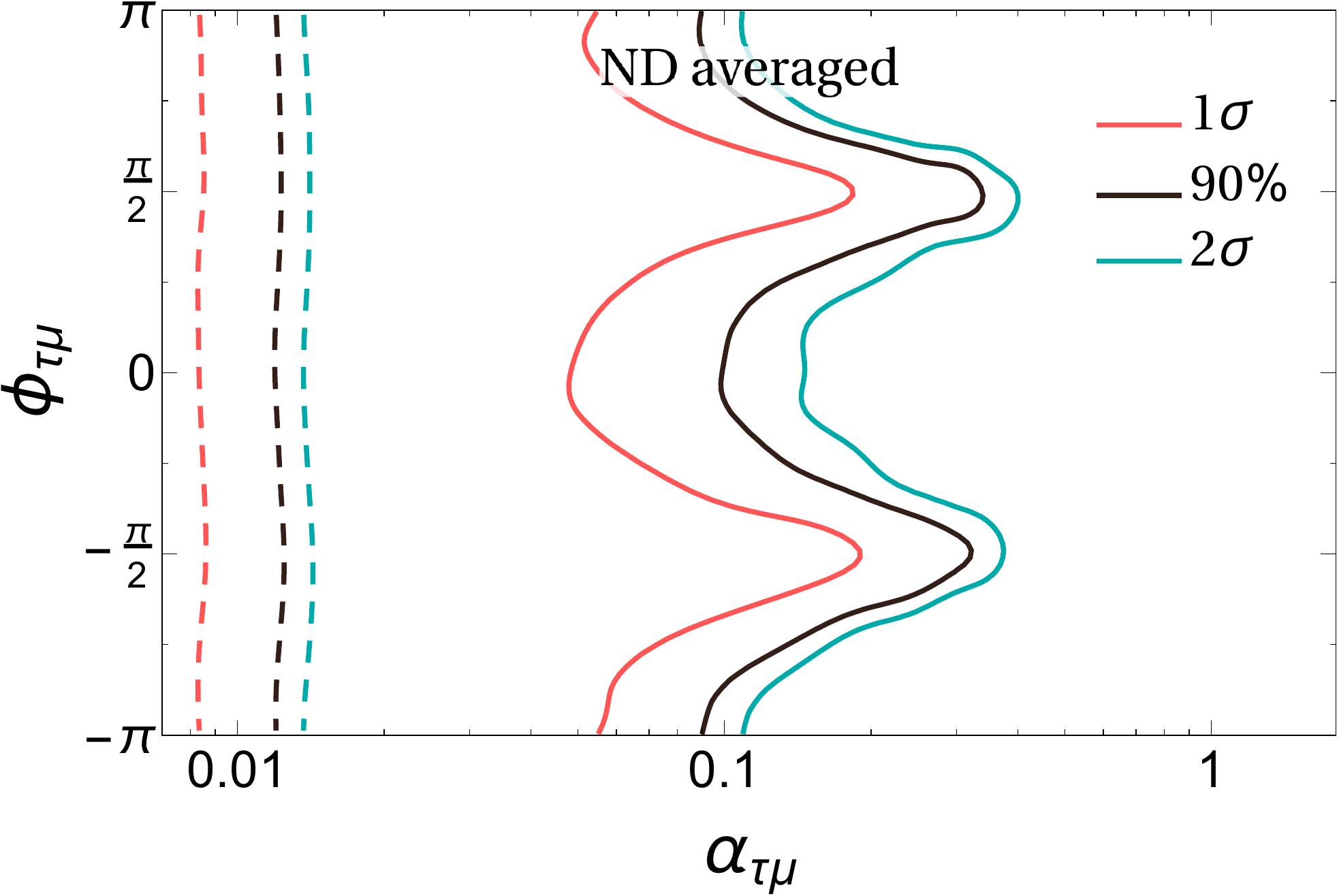} 
 \includegraphics[width=0.48\columnwidth]{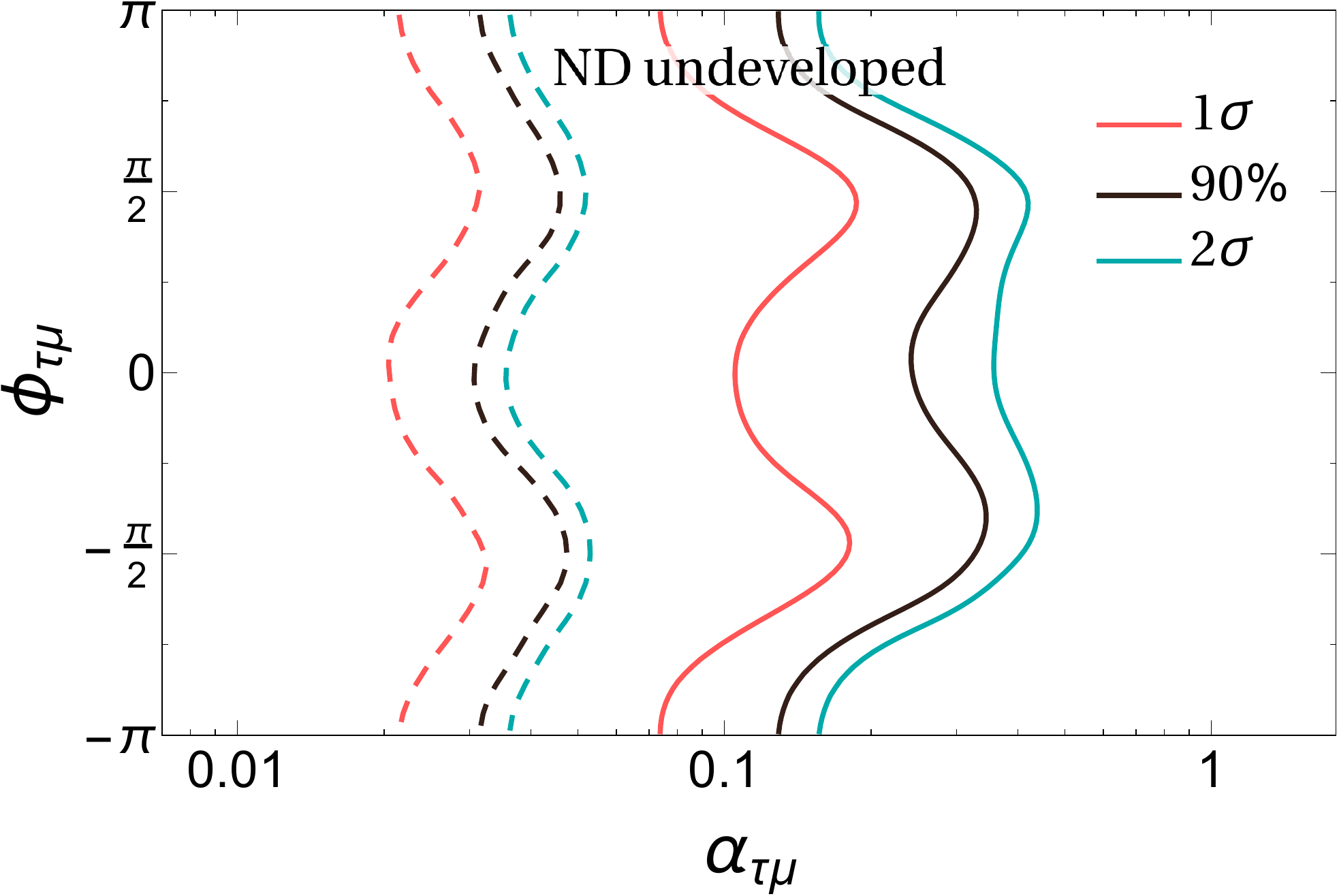}
 \begin{center}
  \caption{Expected frequentist allowed regions at $1 \sigma$, $90\%$ and $2 \sigma$ for DUNE. The sensitivity to the $\alpha_{\beta\delta}$ parameters not shown here can be found in~\cite{Blennow:2016jkn}.\label{fig:plots}}
 \end{center}
\end{figure}
Figure~\ref{fig:plots} shows the expected sensitivities to the new physics parameters. These have been obtained by 
assuming that the true values of all $\alpha$ entries are zero to obtain the corresponding expected number of events, 
and fitting for the corresponding parameters while marginalizing over all other standard and new physics parameters. 
The resulting frequentist allowed regions are shown at  at $1 \sigma$, $90\%$, and $2 \sigma$~C.L. The case in which 
DUNE data is complemented by our present prior constraints on the sterile neutrino mixing (middle and right columns of 
Tab.~\ref{tab:bounds} for the ND averaged and undeveloped scenarios respectively) is depicted with dashed lines while
the for the solid lines no prior constraints were included. Notice that the sensitivities obtained for all parameters 
fall at least one order of magnitude short of the current bounds on the NU from heavy neutrino scenario presented in 
Tab.~\ref{tab:bounds} (left column). It is remarkable that the sensitivity to the real part of 
$\alpha_{\tau\mu}$ improves for the ND undeveloped scenario through the combination of DUNE data and the present
priors with respect to both datasets independently, showing an interesting synergy between data sets.
Another conclusion that can be drawn from Fig.~\ref{fig:plots} is 
that the sensitivities to the diagonal parameters $\alpha_{ee}$ and $\alpha_{\mu\mu}$ are significantly stronger 
for the {ND undeveloped} (right panels) as compared to the {ND averaged} scenario (left panels). This was to be expected 
since the source and detection effects that provide a leading order sensitivity to the diagonal parameters are totally or 
partially cancelled once the normalization of Eq.~\eqref{eq:experimentalprobability} is included 
(see Eq.~(\ref{eq:cancellation})). In the disappearance channel both effects cancel in the ratio, while for the 
appearance channel there is a partial cancellation that only allows the experiment to be sensitive to the 
combination $\alpha_{ee}-\alpha_{\mu \mu}$. This leads to a pronounced correlation among 
$\alpha_{ee}$ and $\alpha_{\mu \mu}$, seen in the upper left panel of Fig.~\ref{fig:plots}. 
In summary, if both near and far detectors are affected by the new physics in the same way 
(as it is the case when the sterile neutrino oscillations are averaged out at both detectors, or in the NU scenario) their 
effects are more difficult to observe since they cannot be disentangled from the flux and cross section determination at 
the near detector. 

%%%%%%%%%%%%%%%%%%%
\section{Conclusions}
\label{sec:concl}
%%%%%%%%%%%%%%%%%%%

We have shown that, when the sterile neutrino oscillations are averaged out 
(and at leading order in the small heavy-active mixing angles) both kinematically accessible sterile neutrinos
and PMNS NU stemming from heavy new physics lead to the same modifications in the neutrino oscillation probabilities. 
However, the present constraints which apply to these two scenarios are very different. Indeed, PMNS NU is bounded at the per 
mille level, or even better for some elements, through precision electroweak and flavour observables, while 
sterile neutrino mixing in the averaged-out regime is allowed at the percent level since it can only be 
probed via oscillation experiments themselves. Thus, no impact in present or near-future 
oscillation facilities from PMNS NU is expected while sterile neutrino mixing could potentially be discovered 
by them if the sterile neutrinos are light enough to be produced at the source. Indeed, our simulations confirm that PMNS NU is beyond the reach of high precision experiments 
such as DUNE, but that sterile neutrino oscillations could manifest in several possible interesting ways. 
Through these simulations the importance of correctly accounting for the impact of the near detector was made evident. 
For instance, a very significant increase in the sensitivity to the new physics parameters 
$\alpha_{ee}$, $\alpha_{\mu\mu}$ was found for the case in which the near 
detector is not affected in the same way as the far detector.

\Acknowledgements
We warmly thank Mattias Blennow, Pilar Coloma, and Enrique Fernandez-Martinez for useful discussion on the preparation of the plenary talk and the poster presented at NuPhys1016. We also acknowledge the HPC-Hydra cluster at IFT.

\end{document}

%% file: econfmacros.tex
\def\beq{\begin{equation}}
\def\eeq#1{\label{#1}\end{equation}}
\def\eeqn{\end{equation}}

\def\beqa{\begin{eqnarray}}
\def\eeqa#1{\label{#1}\end{eqnarray}}
\def\eeqan{\end{eqnarray}}

\let\bar=\overbar

\def\Dslash{\not{\hbox{\kern-4pt $D$}}}
\def\dslash{\not{\hbox{\kern-2pt $\del$}}}

\def\msb{{\bar{\ssstyle M \kern -1pt S}}}